\documentclass{article}

\usepackage{arxiv}

\usepackage[utf8]{inputenc} % allow utf-8 input
\usepackage[T1]{fontenc}    % use 8-bit T1 fonts
\usepackage{hyperref}       % hyperlinks
\usepackage{url}            % simple URL typesetting
\usepackage{booktabs}       % professional-quality tables
\usepackage{amsfonts}       % blackboard math symbols
\usepackage{nicefrac}       % compact symbols for 1/2, etc.
\usepackage{microtype}      % microtypography
\usepackage{lipsum}
\usepackage{graphicx,verbatim,float,subfigure,graphicx,color,siunitx}
\usepackage{amsmath}
\graphicspath{ {./images/} }

\usepackage{amssymb}
\usepackage{svg}
\usepackage{multirow}  %Adds multi-row in tables
\usepackage[export]{adjustbox}
\usepackage{soul}
\usepackage{textcomp}
\usepackage{multirow}
\usepackage{enumitem}
\usepackage[sort&compress,numbers]{natbib}

\title{Moisture-Driven Morphology Changes in the Thermal and Dielectric Properties of TPU-Based Syntactic Foams}

\author{
 Sabarinathan P Subramaniyan\\
  Dept. of Mechanical Engineering \\
  University of Wisconsin-Madison \\
  Madison, WI 53706  \vspace{0.05in} \\
  \And
  Partha Pratim Das\\
  Dept. of Mechanical and Aerospace Engineering \\
  University of Texas at Arlington \\
  Arlington, TX   \vspace{0.05in} \\
  \And
  Rassel Raihan\\
  Dept. of Mechanical and Aerospace Engineering \\
  University of Texas at Arlington \\
  Arlington, TX  \vspace{0.05in} \\
  \And
  Pavana Prabhakar$^*$ \\
  Dept. of Mechanical Engineering \\
  Dept. of Civil \& Env. Engineering \\
  University of Wisconsin-Madison \\
  Madison, WI 53706  \vspace{0.05in} \\
  \texttt{$^*$pavana.prabhakar@wisc.edu}
}

\begin{document}
\maketitle

\newcommand{\SPS}[1]{\textcolor{red}{\bf{Sabari: #1}}}
%\newcommand{\VD}[1]{\textcolor{blue}{\bf{ #1}}}
%\newcommand{\pavana}[1]{\textcolor{blue}{\bf{ #1}}}
%----------------------------------------------------------------------------------------
%	TITLE SECTION
%----------------------------------------------------------------------------------------

\begin{abstract}
Syntactic foams are a promising candidate for applications in marine and oil and gas industries in underwater cables and pipelines due to their excellent insulation properties. The effective transmission of electrical energy through cables requires insulation materials with a low loss factor and low dielectric constant. Similarly, in transporting fluid through pipelines, thermal insulation is crucial. However, both applications are susceptible to potential environmental degradation from moisture exposure, which can significantly impact the material's properties. This study addresses the knowledge gap by examining the implications of prolonged moisture exposure on TPU and TPU-derived syntactic foam via various multi-scale materials characterization methods. The research focuses on a flexible syntactic foam created using selective laser sintering and thermoplastic polyurethane elastomer (TPU) reinforced with glass microballoons (GMB). The study specifically explores the impact of moisture exposure duration and GMB volume fraction on microphase morphological changes, their associated mechanisms, and their influence on thermal transport and dielectric properties.
\end{abstract}

% keywords can be removed

%----------------------------------------------------------------------------------------
%	ARTICLE CONTENTS
%----------------------------------------------------------------------------------------

\section{Introduction}\label{intro}
Block copolymers known as thermoplastic polyurethane elastomers (TPUs) are composed of alternating hard and soft segments, with the soft segments typically derived from polyester or polyether (polyol) and the hard segments produced by combining a chain extender with an isocyanate. By varying the formulation and constitution, as well as incorporating additives, the mechanical properties of TPUs, including modulus, strength, hardness, damping, and tribological performance, can be customized. TPUs are incredibly versatile, with widespread use in various applications such as sealings, hoses, shoe soles, cable sheaths, films, foams, and automotive interiors. They exhibit the processing abilities of plastics and the high elasticity of rubber, making them a sought-after material in the industry. Moreover, TPUs are well-known for their low-temperature flexibility, excellent abrasion resistance, good processing characteristics, and biocompatibility, making them ideal for use in diverse fields, including transportation, construction, and biomedical materials\cite{Aurilia2011,Bruckmoser2014,Deng1994,Osswald2012,Xu2021}.

Considerable research has been carried out to investigate the impact of hard-to-soft segment ratio, hard segment type, chain extender type, and segment length on various properties of thermoplastic polyurethane elastomer (TPU). Hydrogen bonding, crystallization behavior, mechanical performance, thermal transport, radiation stability, and dielectric properties of TPU have been extensively studied\cite{Xiu1992,Xiang2017,Walo2014,Petcharoen2013,Li2019,Jomaa2015}. Numerous studies have been conducted on the impact of moisture on TPU, with a notable emphasis on mechanical degradation\cite{Yang2006,Boubakri2009,Boubakri2010, Puentes-Parodi2019,Bardin2020,Xu2021, mishra2015long,choi2023degradation}. Despite the large volume of research that has been conducted on TPU, a significant knowledge gap remains regarding the long-term effects of moisture-induced microphase morphology changes caused by hydrolysis, and specifically how these changes affect  thermal transport and dielectric properties. Our current study aims to bridge this gap by establishing a correlation between microphase morphology and its properties.

Syntactic foams are a type of closed-cell foam that is composed of hollow microspheres that are embedded within a matrix material. The microspheres themselves are typically constructed from materials such as glass, cenosphere (a waste material produced by the combustion of fly ash), or metal, while the matrix materials are typically composed of either polymers, ceramics, or metals. Among the various types of syntactic foams, polymer-based foams are particularly popular in a range of different industries, including the marine, automotive, aerospace, and electrical sectors. This is largely due to the fact that they offer numerous desirable properties, such as high strength, low density, high buoyancy, low thermal expansion, vibrational damping, and acoustic, thermal, and electrical insulation. These properties make them ideal for a wide range of applications, including energy-absorbing cores in sandwich composites, radome material, acoustic insulation for underwater sonar devices, thermal insulation, and electrical cable insulation. Importantly, the properties of syntactic foams can be customized by altering the matrix type, the volume fraction of the hollow microspheres, and the thickness of the hollow microspheres' walls\cite{Gupta2014,Afolabi2020,Shahapurkar2018,GARCIA2018,PRABHAKAR2022-CompB}. In the present work, thermoplastic polyurethane elastomer was used as the matrix material in conjunction with soda-lime borosilicate-based glass microballoon reinforcements.

Elastomeric and rubber-based syntactic foams have been the subject of significant advancements in recent years, with researchers developing materials that exhibit remarkable versatility across a range of industries. Notably, these foams have been employed in the design of shoe soles, pneumatic tires, wires, and cable compounds. Furthermore, recent research has demonstrated the immense potential of thermoplastic polyolefin elastomers-based syntactic foam for buoyancy and coaxial cable insulation\cite{amos2015hollow}. Tripathi et al.\cite{tripathi2022flexible} developed a fire-protective clothing material comprising flexible silicone hollow glass microballoons syntactic foam. Their investigation revealed that applying syntactic foam coating on the glass fabric resulted in a notable improvement, with a second-degree burn time of 10 seconds, compared to the 6.4 seconds observed for the glass fabric alone. In the current study, Thermoplastic Polyurethane Elastomer (TPU) reinforced with Glass MicroBalloons (GMB) was manufactured using Selective Laser Sintering (SLS). 

In our prior study \cite{SUBRAMANIYAN2023110547}, we conducted a deep dive into understanding the role of GMB reinforcement and moisture aging on the degradation mechanisms in the viscoelastic properties of TPU-Based Syntactic Foams produced using the SLS process \cite{Tewani2022}.
The existing literature on TPU-based composites produced via SLS has predominantly focused on boosting thermal and electrical conduction properties\cite{Zhou2022,Zhang2022,Gan2019,Yuan2018,Li2017,Shen2023,zhang2022combination} — consequently, a gap in the realm of thermal and electrical insulation warrants further investigation.

To that end, the current work aims to investigate the effect of moisture, GMB volume fractions, and temperature on the material's thermal transport and dielectric properties. The chemical changes resulting from moisture aging were also studied and correlated with the observed material properties.

%%%%%%%%%%%%%%%%%%%%%%%%%%%%%%%%%%%%%%%%%%%%%%%%%%%%%%%%%%%%%%%%%%%%%%%%%%%%%%%%%%%%%%%%%%%%%%%%%%%%%%%%%
\section{Motivation}
The primary aim of this study is to explicate the impact of internal structure and moisture-triggered deterioration on the functional characteristics of TPU-based syntactic foams fabricated through additive manufacturing. Furthermore, this paper investigates how prolonged exposure to moisture alters the microphase morphology and chemical composition of TPU- and TPU-based syntactic foams and the consequential effects on their thermal transport and dielectric properties.

\begin{figure}[h!]
 \centering
  \includegraphics[width=14cm]{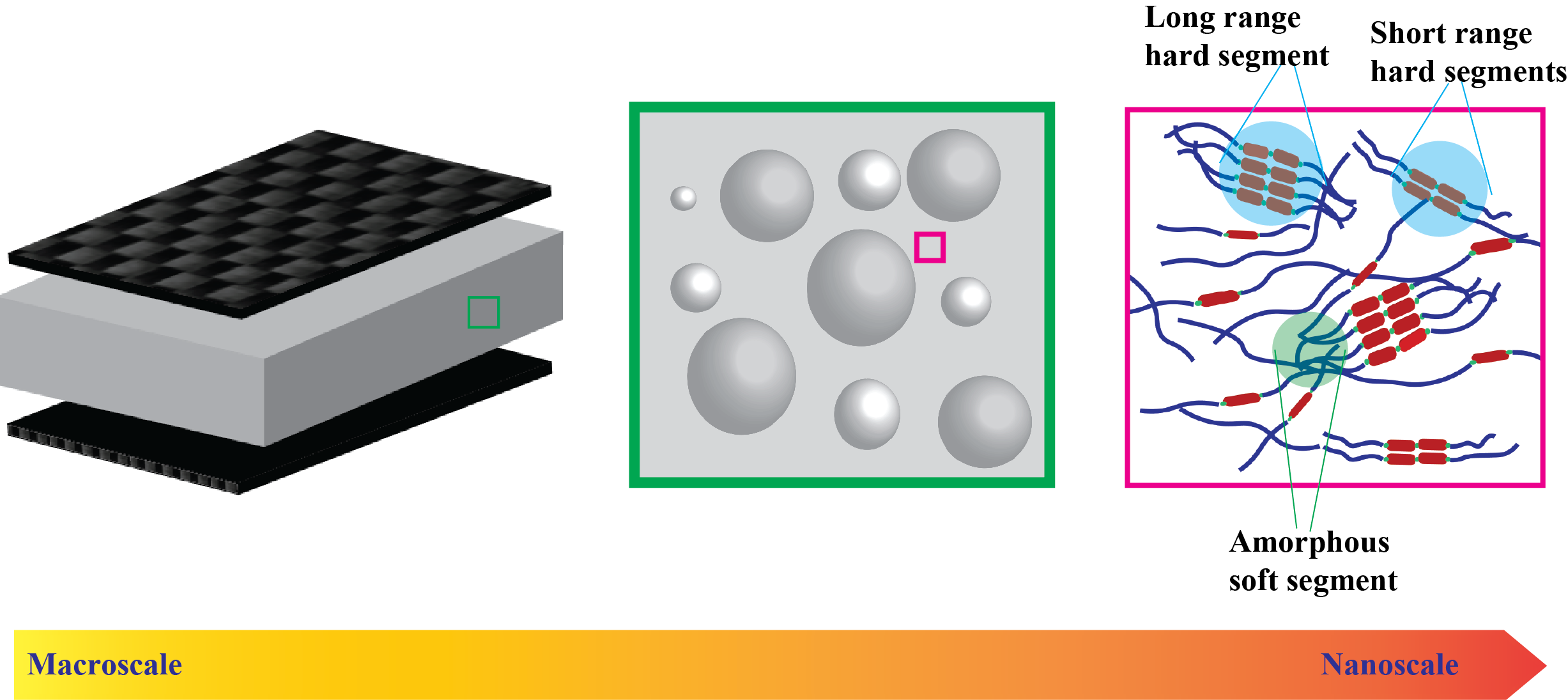}
  \caption{Illustration of hollow glass microballoon (GMB) reinforced Thermoplastic Polyurethane (TPU) elastomer at different length scale}\label{fig:material}
\end{figure}

\section{Methodology}\label{method}

\section{Materials}

Fig~\ref{fig:material} illustrates the multiscale material hierarchy of TPU-based syntactic foam. At the macroscale level, this foam is a lightweight structure that can be customized into different architectures using additive manufacturing techniques, allowing for control over its macroscale property \cite{Tewani2022}. The microscale level of this foam consists of glass micro balloons that are reinforced into a TPU matrix, with the volume fraction being adjustable to achieve the desired density and property of interest. At the nanoscale level, TPU is a block copolymer composed of hard and soft segments. The hard segments are obtained by reacting diisocyanates with a diol or diamine chain extender, while the soft segments are made from amorphous polyester or polyether. The thermodynamic incompatibility between the soft and hard segments might cause macrophase separation, a common occurrence in polymer blends, and can affect physical properties. However, covalent bonds between hard and soft segments in TPU prevent the formation of macrophase separation, creating different microphase-separated morphology. Although these structures are observed at the nanoscale, we will use the older terminology of microphase morphology. Modifications in the microphase morphology of TPU play a vital role in determining the functional properties of the material \cite{yilgor2015critical,cheng2022review}.

\subsection{Manufacturing and conditioning}
In this study, we examine syntactic foams that are fabricated at the University of Wisconsin-Madison from a Thermoplastic polyurethane elastomer matrix (FLEXA Grey - Sinterit) and hollow micro-glass balloon fillers (K20 - 3M), which are produced through the process of selective laser sintering. The composite powder preparation process entails blending TPU with GMB at varying volume fractions (20\% and 40\%) using an automated mixer for 5 minutes at 30 V and 3 minutes at 70 V to ensure adequate dispersion of GMB in TPU. The mixed powder is then introduced into the SLS equipment (Lisa-pro) for sintering, with an input laser power ratio of 1.5 and layer height of 0.075 mm. The input parameters are determined based on a parametric study by Tewani et al.\cite{Tewani2022}. Pristine TPU and TPU with 20\% and 40\% GMBs are manufactured and immersed in de-ionized water at 23 \textdegree C. Then, the samples are removed and desorbed at 50 \textdegree C in an oven for 24 hours to remove the free water in the samples. The primary focus of this work is to understand the underlying mechanism of bound water's influence on thermal and dielectric properties.

\subsection{Thermal stability characterization}
Differential Scanning Calorimetry (DSC), using TA equipment QA 200, is carried out at the University of Wisconsin-Madison to comprehend the effects of moisture and GMB volume fraction on the thermal stability of TPU-based syntactic foam. We employ a sample weight of 4–8 mg in a controlled inert atmosphere to ignore the complexity of oxygen-induced breakdown. By heating from -50\textdegree C to 225 \textdegree C and cooling back down to -50 \textdegree C, we undertake dynamic analysis at a constant temperature ramp rate of 10 \textdegree C/min. Calculating melting enthalpy ($\Delta H_{T}$) involves using the second heating curve.

\subsection{Thermal transport characterization}
\begin{figure}[h!]
\centering
\subfigure[]{
\includegraphics[width=5cm]{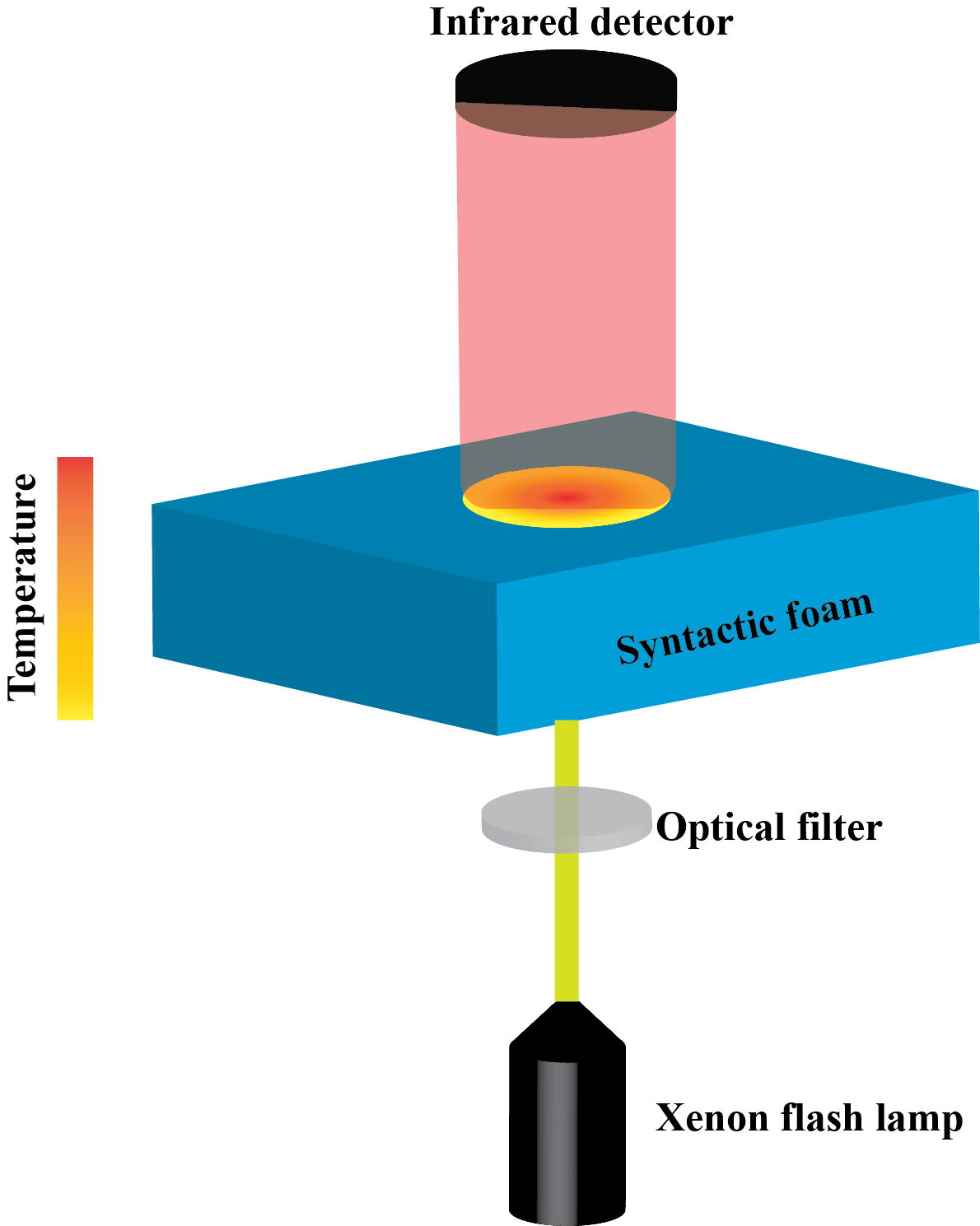}\label{LFA_procedure}
}
\hspace{0.2in}
\centering
\subfigure[]{
\includegraphics[width=8cm]{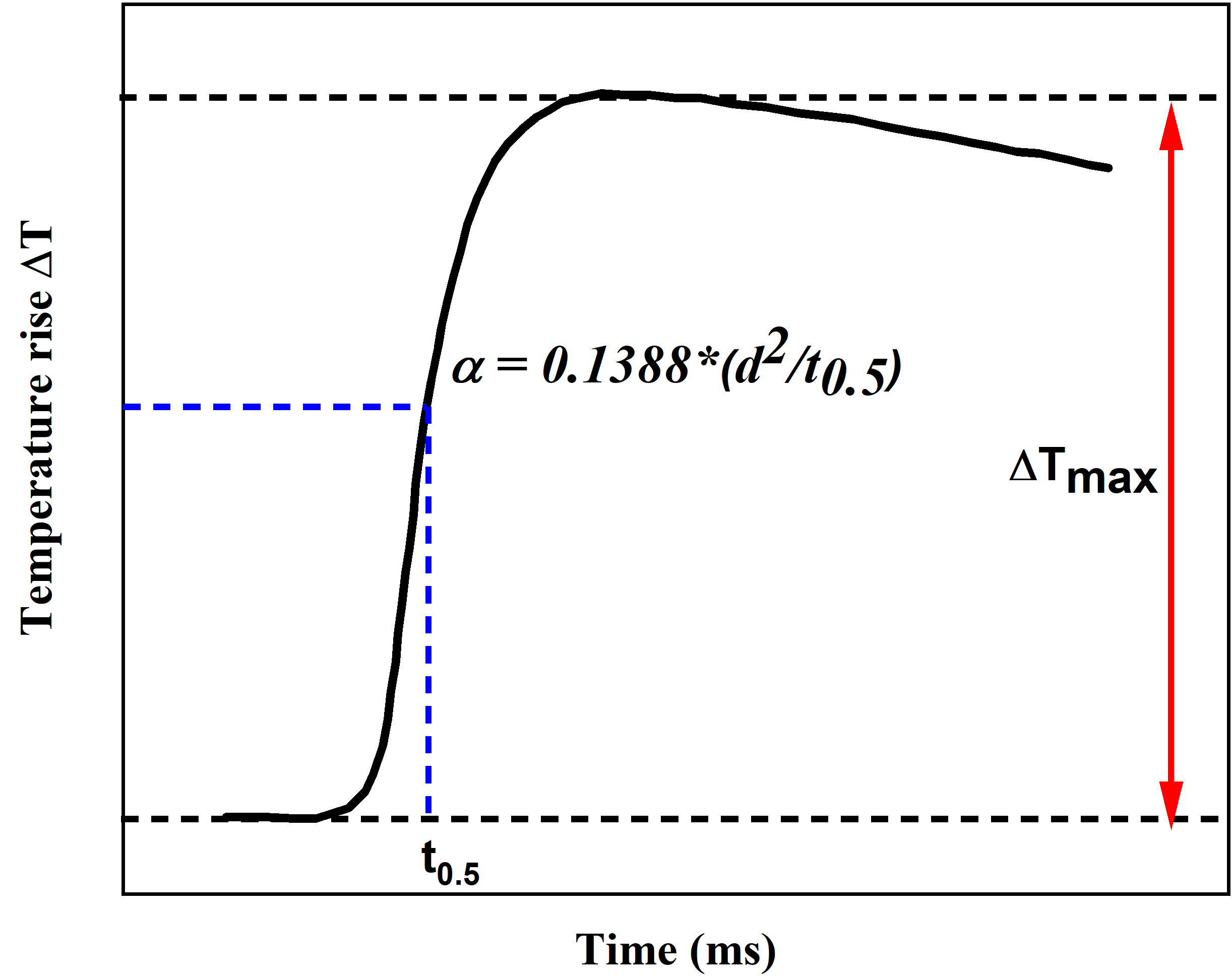} \label{LFA_measure}
}
\caption{(a) Schematic illustration of laser flash analysis and (b) Thermal diffusivity measurement using Time vs. Temperature rise plot}
\end{figure}

Laser Flash Analysis (LFA) is a widely used thermal analysis technique to measure thermal transport properties. It involves using a high-intensity laser to generate a short pulse of heat that is applied to a small spot on the rear surface of a sample. The resulting temperature rise on the front surface is measured using a detector as shown in Fig~\ref{LFA_procedure}. Later, the thermal diffusivity is calculated from the temperature rise signal shown in Fig~\ref{LFA_measure}. In this work, the thermal diffusivity of TPU and TPU-based syntactic foam is measured using NETZSCH LFA 447 at the University of Wisconsin-Madison over a temperature range from 25°C to 100°C. For the purposes of this investigation, a specimen with dimensions of 10 x 10 x 1 $mm^3$ is chosen, and the half rise time is measured using the Radiation+pulse model in Netzsch proteus analysis software.

\begin{equation}
    \alpha = 0.1388*\frac{d^2}{t_{0.5}}
\end{equation}
where $\alpha$, $d$, and $t_{0.5}$ refer to thermal diffusivity in $mm^2/s$, sample thickness in mm, and half raise time as shown in Fig~\ref{LFA_measure}. The specific heat of the samples is measured using DSC, and the measured density of syntactic foams is input into Equation~\ref{conduc} to obtain the thermal conductivity $\kappa$
\begin{equation}\label{conduc}
    \kappa = \alpha*\rho*C_p
\end{equation}

\subsection{Dielectric characterization}
The dielectric properties of the syntactic foam are measured at the University of Texas at Arlington using Novocontrol Broadband Dielectric Spectroscopy assisted with an alpha analyzer to measure the complex dielectric and impedance of syntactic foams as a function of frequency. The dielectric properties are measured at room temperature by utilizing the parallel plate capacitor, where 1000mV of voltage is applied across a sample size of 30 mm diameter with 1 mm thickness at frequencies ranging from 10 Hz to $10^6$ Hz. The frequency range under consideration encompasses valuable information regarding molecular and dipolar fluctuations. Furthermore, charge transport and polarization effects, which manifest at both the inner and outer boundaries, can be effectively captured through dielectric properties\cite{raihan2014dielectric}. The dielectric values are obtained by analyzing the phase shift between the applied voltage and the current response. The real and imaginary part of complex dielectric is called dielectric constant and dielectric loss, respectively. tan$\phi$ is dielectric loss tangent.

%%%%%%%%%%%%%%%%%%%%%%%%%%%%%%%%%%%%%%%%%%%%%%%%%%%%%%%%%%%%%%%%%%%%%%%%%%%%%%%%%%%%%%%%%%%%%%%%%%%%%%%%%

\section{Results}\label{res}

This study investigated the influence of moisture on TPU (thermoplastic polyurethane) and TPU-based syntactic foams. Section~\ref{DSC-thermal_characterization} uses DSC to characterize the material's thermal behavior. The study involves deconvoluting the endothermic peaks and evaluating melting enthalpies to determine different microphase morphology changes. Thermal transport changes induced by moisture in TPU-based syntactic foams are examined in Section~\ref{thermal_transport_characterization}, while Section~\ref{dielectric_characterization} focuses on analyzing dielectric behavior changes resulting from moisture exposure. Overall, this work provides a comprehensive investigation into the influence of moisture on TPU and TPU-based syntactic foams, focusing on thermal and dielectric behavior changes.

%\begin{comment}
\begin{figure}[h!]
 \centering
  \includegraphics[width=12cm]{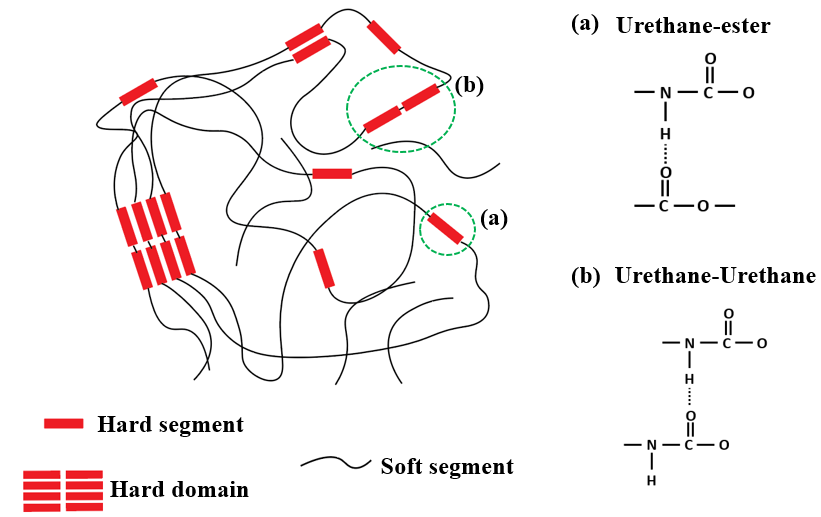}
  \caption{Schematic representation of  hydrogen bonding in thermoplastic polyurethane elastomer}\label{Bond_type}
\end{figure}
%\end{comment}

\subsection{Influence of moisture and GMB volume fraction on the thermal stability of syntactic foam}\label{DSC-thermal_characterization}

\begin{figure}[h!]
\centering
\subfigure[]{
\includegraphics[width=6.8cm]{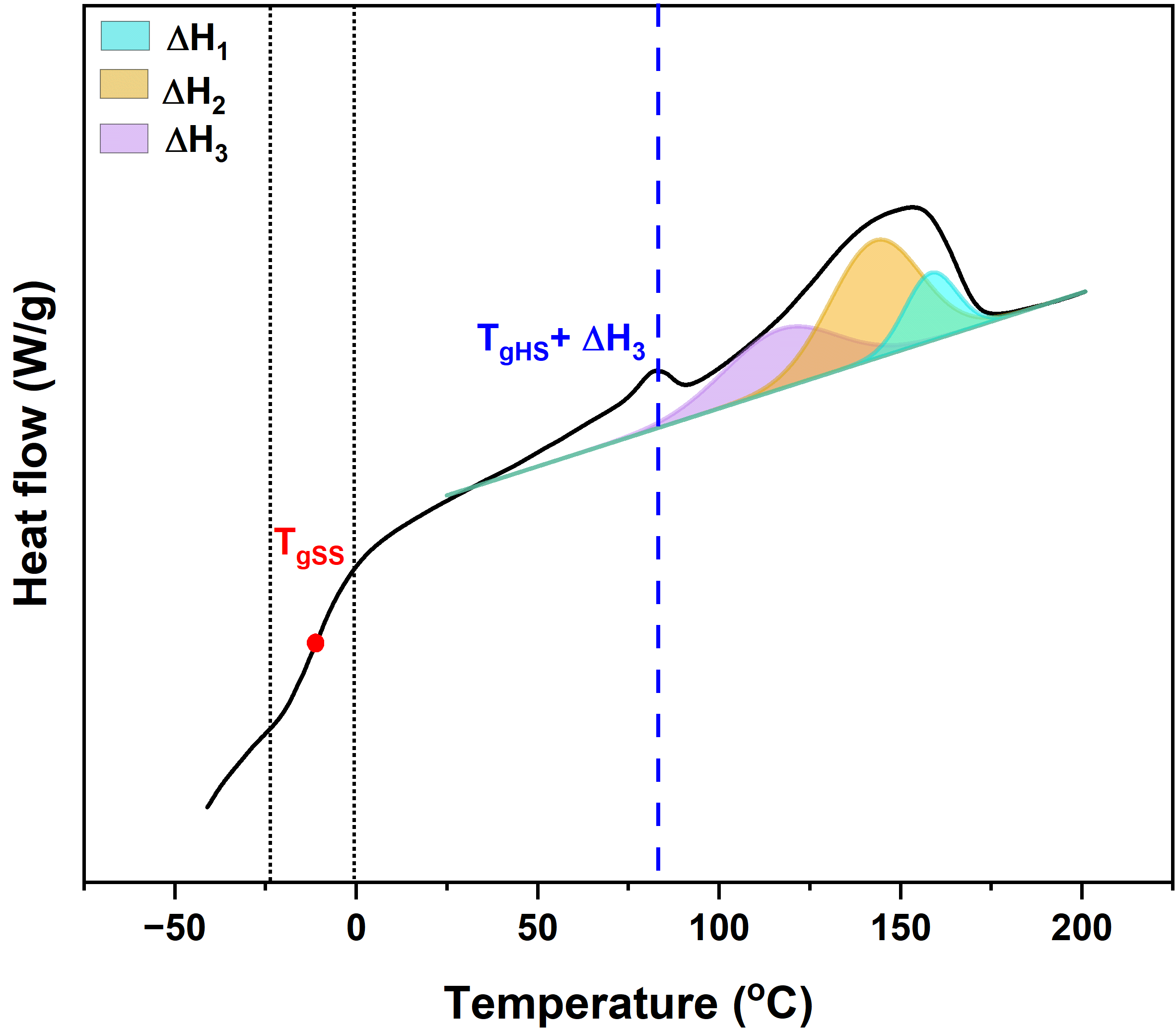}\label{DSC_peaks}
}
\hspace{0.2in}
\centering
\subfigure[]{
\includegraphics[width=7.25cm]{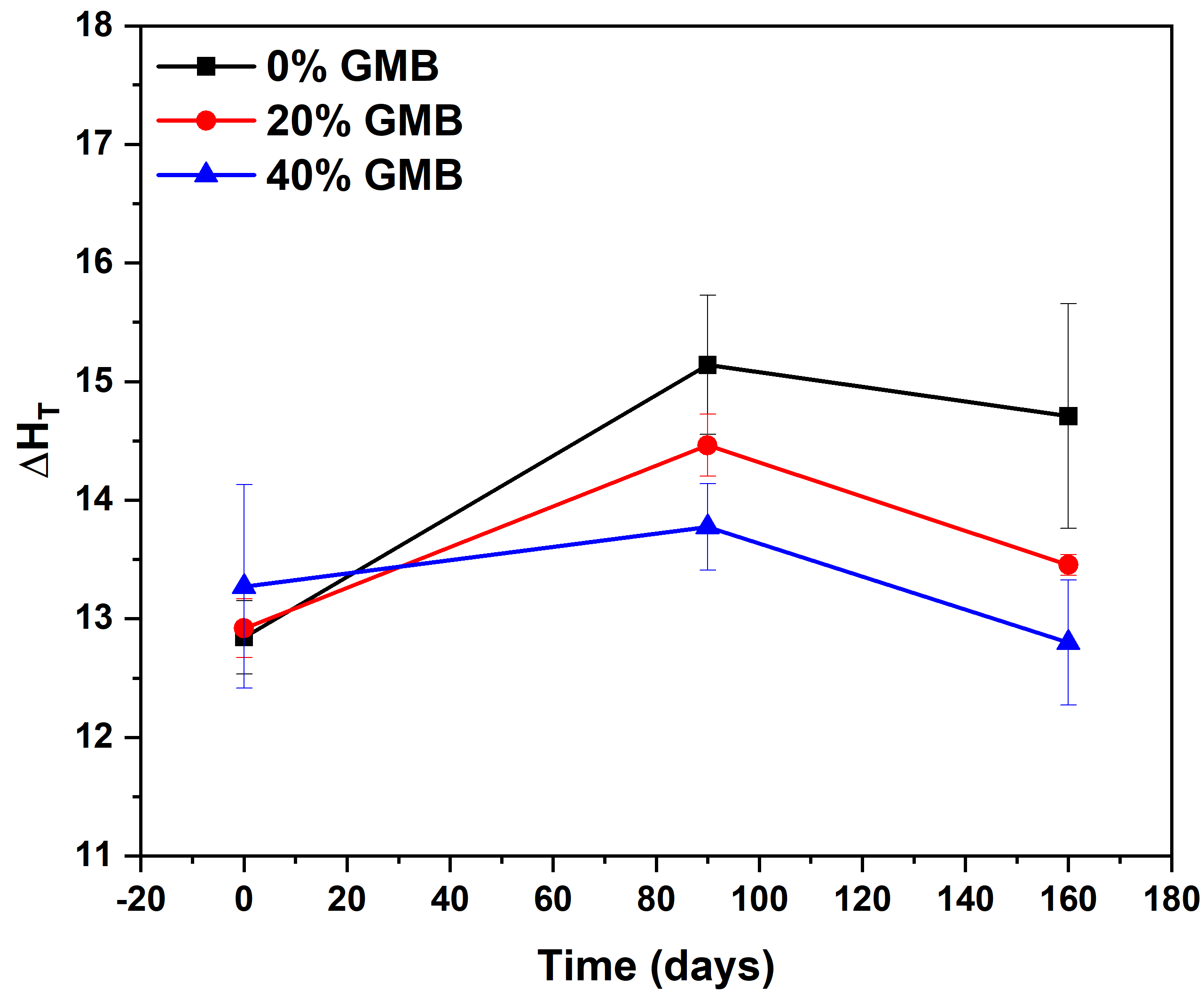}\label{enthalpy}
}
\caption{(a)The endothermic heating cycle of thermoplastic polyurethane reveals hidden peaks that distinctly outline various hard segment morphologies.  (b) Total melting enthalpy for different volume fractions of GMB at different moisture exposure times.}
\end{figure}

Differential scanning calorimetry (DSC) has been shown to be an effective means of capturing moisture-induced microphase morphology changes in TPU and TPU-based syntactic foam. Typically, the second heating curve of the samples is used for analysis, as the first heating curve may reflect the material's thermal history. The glass transition temperature of the soft segment ($T_gSS$) is attributed to a weak transition in the sub-zero degree temperature range, which is observed in the endothermic cycle (as shown in Fig~\ref{DSC_peaks}). The hydrogen bonds between hard and soft segments, as well as between hard segments, break and dissociate upon heating, followed by melting of hard segments involving different transitions and multiple endothermic peaks, as illustrated in Fig~\ref{DSC_peaks}. As the sample is heated, a weak transition accompanied by a small endothermic peak appears at around 80 \textdegree C, primarily due to the glass transition temperature of the hard segment ($T_gHS$) and the annealing endotherm, which appears at 20-30 \textdegree C higher than the annealing temperature\cite{Saiani2007} (50 \textdegree C in this paper). This annealing endotherm is often associated with the onset endothermic peak of short-ordered crystallites of the hard segment\cite{Kong2019,Puentes-Parodi2019,Verbelen2017}.
 
Upon further heating, a broad transition peak was observed in Fig~\ref{DSC_peaks} that encompasses multiple concealed peaks. These concealed peaks are associated with the long-range hard segment possessing distinct morphologies. The ultimate melting enthalpy peaks correspond to the long-range-ordered crystallites of the hard segment that necessitate more energy to break down and melt than the preceding peaks of melting enthalpy. The comprehensive melting enthalpy ($\Delta H_{T}$) is the amalgamation of the areas beneath $\Delta H_{1}$, $\Delta H_{2}$, and $\Delta H_{3}$, which pertain to the melting enthalpy of micro-crystalline hard domains with highly ordered long-range hard segments, disordered long-range hard segments, and para-crystalline with short-range hard segments, respectively \cite{razeghi2018tpu}.

\begin{figure}[h!]
 \centering
  \includegraphics[width=14cm]{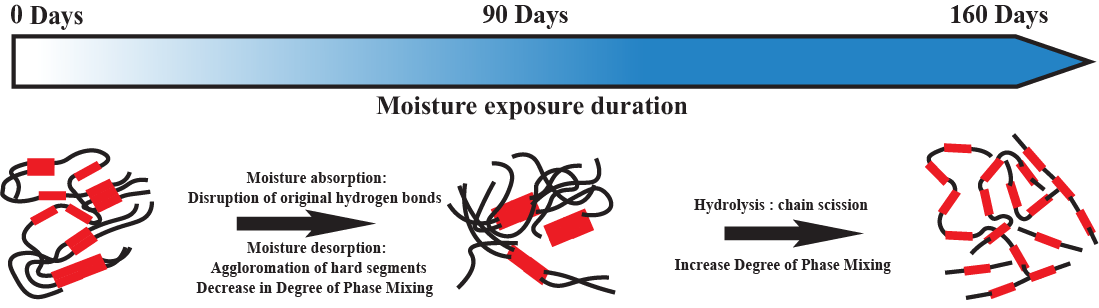}
  \caption{Illustration of moisture-induced mechanisms in TPU elastomer at different times of exposure}\label{fig:mechanism}
\end{figure}

Upon comparison of the $\Delta H_{T}$ values depicted in Fig~\ref{enthalpy} for pristine samples (unaged) with varying GMB volume fractions, it is evident that the TPU with 40\% GMB exhibits the highest value, followed by TPU with 20\% GMB, and then the Neat TPU. The reason behind this observation can be attributed to the presence of GMBs in the powder mix that can potentially affect laser processing in SLS. The dissimilar optical and thermal properties of GMBs and TPU can lead to non-homogeneous heat being supplied by the laser to melt the TPU and GMB powder mixture. This, in turn, results in more energy being supplied to the TPU within the GMB mixture, thereby increasing the different morphologies of hard phases in the material. The total enthalpy increased significantly for neat TPU after 90 days of moisture exposure (Fig~\ref{enthalpy}), whereas TPU with 20\% GMB showed a slightly lower increase and TPU with 40\% GMB exhibited only a marginal increase. This can be attributed to the fact that the water molecules diffused into the material and attacked the original hydrogen bond (NH---C=O bond) with higher fracture energy, resulting in weak hydrogen bonds with free carbonyl groups or amine groups\cite{Xu2021}. After desorbing the material at 50 °C for 24 hours, the hard segments dissociated, reorganized, and formed different morphologies of crystallites. At 90 days, the polyester was not affected by the water due to hydrolysis-induced chain scission occurring much later. On the other hand, at 160 days of moisture exposure, hydrolysis of the soft segments dominated, causing chain scission of the long-chain polyol to occur at an exponential rate\cite{Bardin2020}. This process produced more short-chain diol which then formed bonds with the urethane hard segments, spreading them homogeneously throughout the multiphase system. Consequently, the total enthalpy decreased for all samples. These mechanisms with exposure time are illustrated in Fig~\ref{fig:mechanism}

\subsection{Influence of moisture and GMB volume fraction on thermal transport properties of syntactic foam}\label{thermal_transport_characterization}

\begin{figure}[h!]
\centering
\subfigure[]{
\includegraphics[width=5cm]{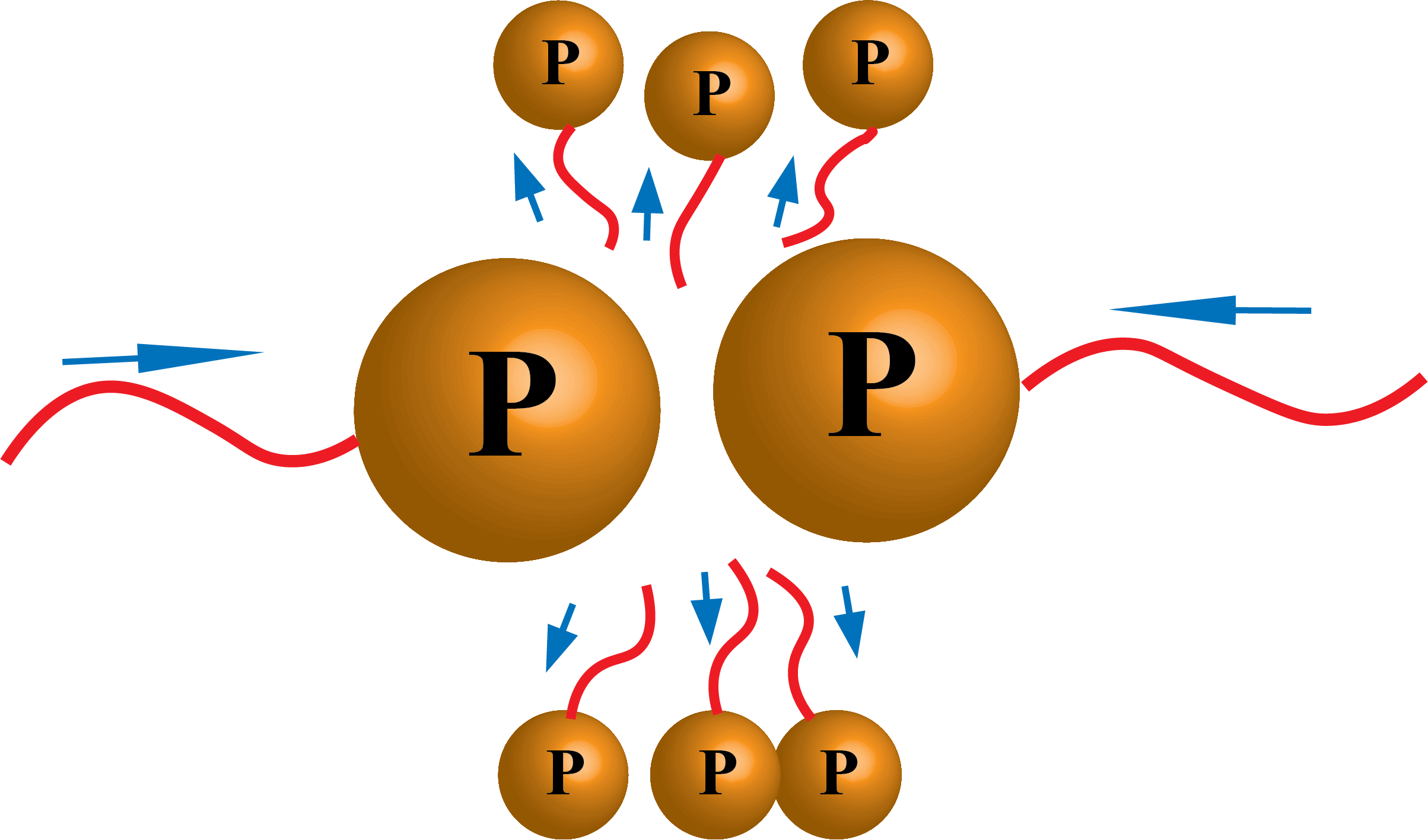}\label{pps}
}
\hspace{0.2in}
\centering
\subfigure[]{
\includegraphics[width=5cm]{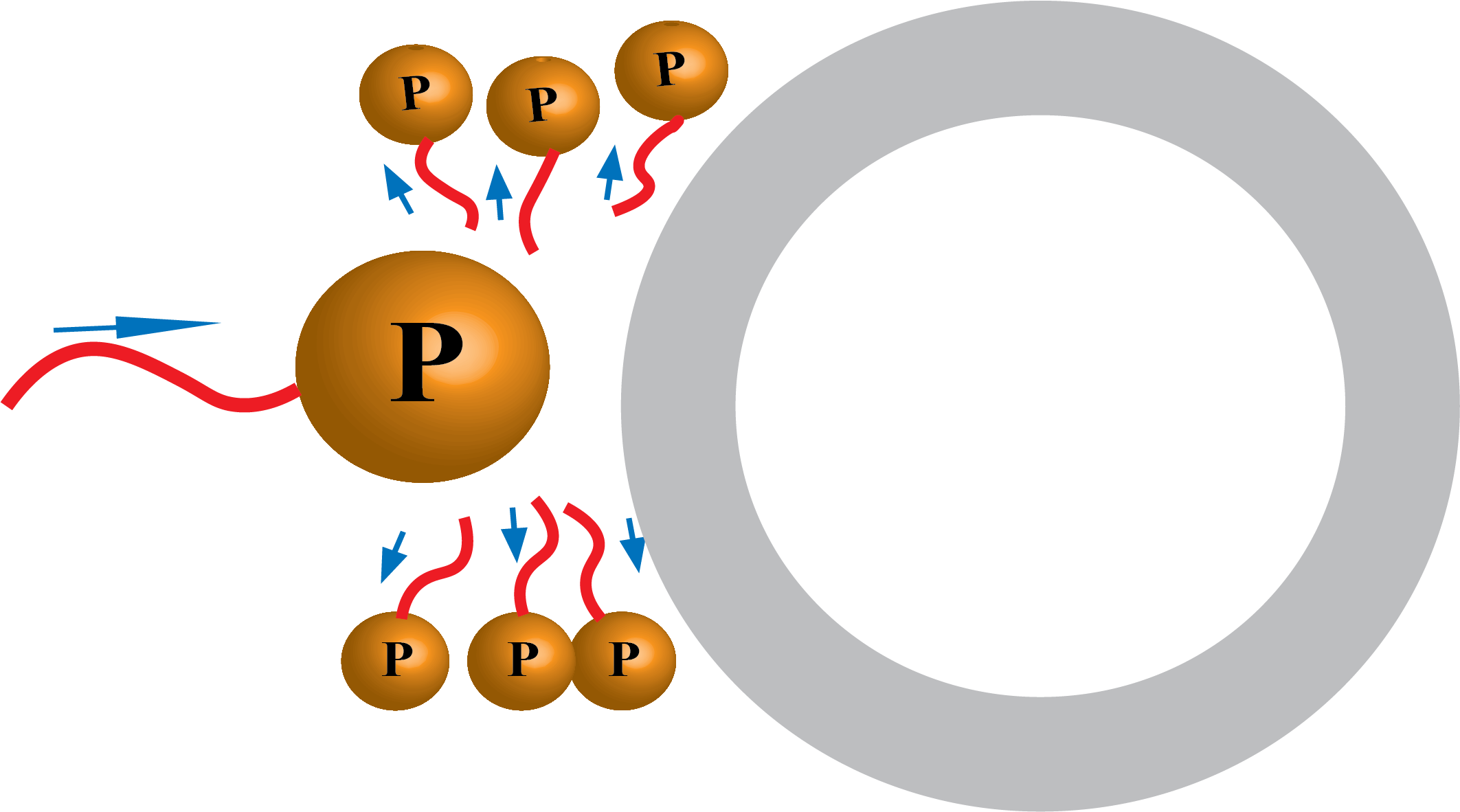}\label{pbs} 
}
\caption{(a) Phonon-Phonon scattering and (b) Phonon-Boundary scattering}
\end{figure}

Phonons are discrete units of energy generated by atomic lattice vibrations, which are responsible for heat transfer in insulating materials such as polymers. The thermal conductivity of polymer and polymer composite materials relies on the average distance Phonons travel before experiencing scattering, known as the mean free path. A higher mean free path results in higher thermal conductivity. Three distinct types of scattering, namely phonon-phonon scattering, phonon-boundary scattering, and phonon-impurity scattering, influence the phonon mean free path. 
Temperature influences the phonon's mean free path, where the wavelength of the dominant phonon decreases with an increase in temperature, which means phonon wavelength is higher at low temperatures. As the temperature increases, the phonon's mean free path decreases due to phonon-phonon scattering as illustrated in Fig.~\ref{pps}, consequently reducing the thermal conductivity. At low temperatures, the long wavelength phonons are not affected by defects or impurities as they are at the atomic scale. Hence, the phonon at low temperatures majorly experiences phonon-boundary scattering as illustrated in Fig.~\ref{pbs}. However, as the temperature increases, the dominant phonon's wavelength becomes comparable to defects and impurities, and phonon-impurity scattering becomes active. In contrast, further increase makes the phonon wavelength to be of size compared to crystal size, so phonon-phonon scattering becomes dominant.\cite{Mehra2018,Guo2020}

With an increase in the volume fraction of GMB in TPU, both the thermal conductivity and specific heat capacity were observed to decrease. For instance, at a temperature of 25°C, it was found that the thermal conductivity of unaged TPU with 20\% GMB and 40\% GMB decreased by 38.42\% and 56.98\%, respectively, when compared with unaged neat TPU. The decrease in thermal conductivity with an increase in the volume fraction of GMB can be attributed to the increase in void contents in the TPU. The thermal conductivity of the K20 GMB utilized in this study is considerably low, measuring only 0.070 W/mK at 25°C. Furthermore, an increase in the volume fraction of GMB also decreases the specific heat capacity. Specifically, the specific heat capacity of a material pertains to the amount of energy required to raise the temperature of 1g of a material by one degree Celsius. Neat TPU exhibits a tendency to absorb more energy in order to increase its temperature, whereas an increase in the volume fraction of GMB reduces the total amount of polymer, subsequently leading to a decrease in the specific heat capacity.

\begin{figure}[h!]
\centering
\subfigure[]{
\includegraphics[width=7cm]{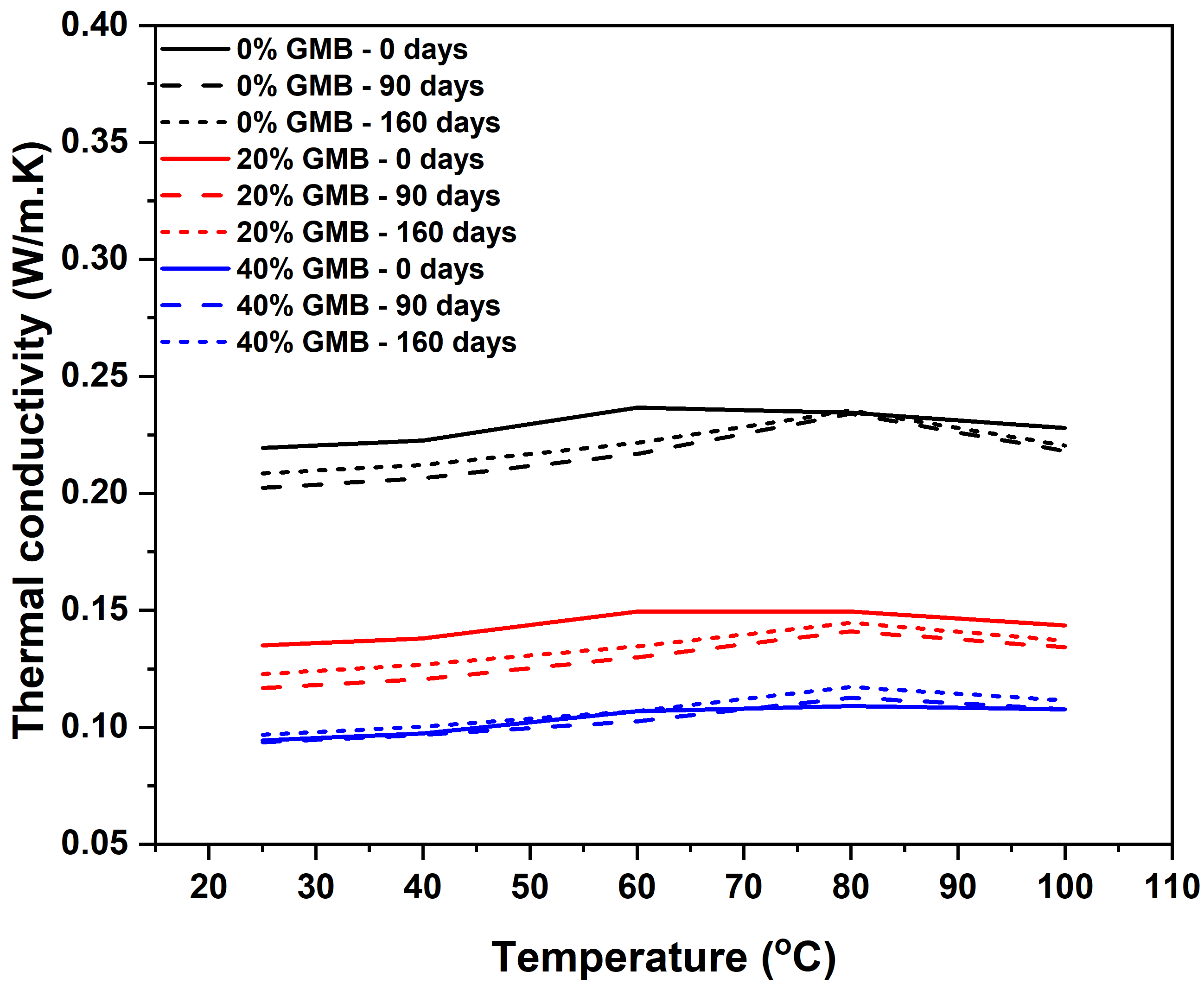}\label{Therm_cond}
}
\hspace{0.1in}
\centering
\subfigure[]{
\includegraphics[width=7cm]{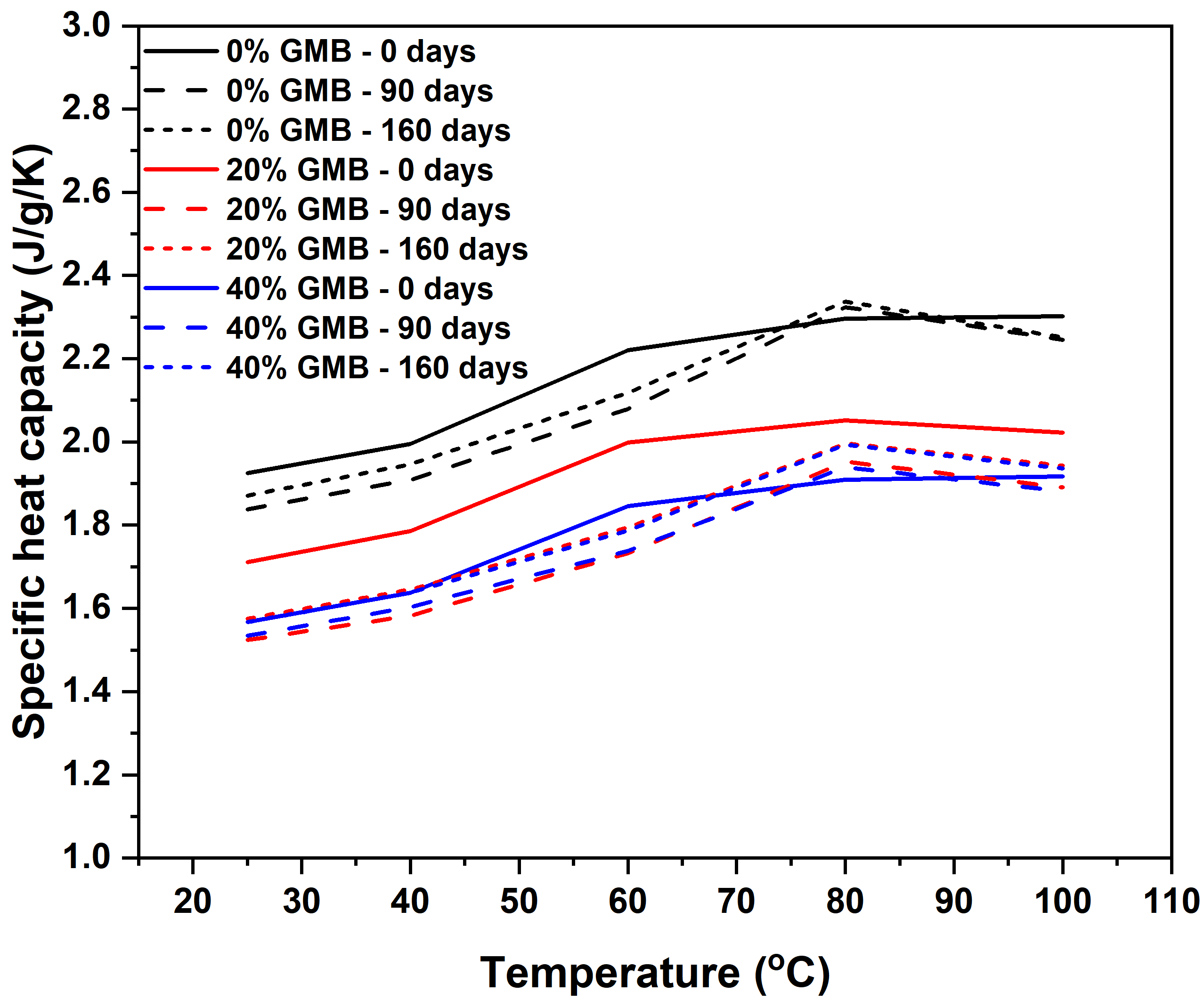}\label{specif_heat}
}
\caption{(a) Thermal conductivity and (b) Specific heat capacity of TPU with different volume fractions of GMB reinforcement and moisture exposure time. }
\end{figure}

The thermal conductivity of the polymer is also dependent on the exposure temperature since the polymer undergoes significant structural changes like glassy below the glass transition temperature, leathery and rubbery above the glass transition temperature, and finally, reaching the terminal region. In this work, TPU is an elastomer with a multiblock structure with different phase-separated morphology that shows complex thermal transport behavior as the temperature changes. So, we opted for a temperature range between 25\textdegree C to 100\textdegree C.
Fig~\ref{Therm_cond} shows that the thermal conductivity of TPU and TPU-based syntactic foams did not change significantly with temperature. However, around 60\textdegree C, there is a slight increase in the thermal conductivity, with marginal reduction until 100\textdegree C. For unaged samples, this rise in thermal conductivity is associated with the melting of short-order crystallites. One main reason for this plateau behavior initially is due to the amorphous nature of the soft segment in TPU. As the temperature increases, the main chain movements will create more microvoids, enhancing the phonon scattering and decreasing the thermal conductivity. At the same time, there is an opposing reaction where the dominant chain movements might increase chain alignments and bring the chain arrays closer, increasing the thermal conductivity and eventually resulting in a plateau over a specific temperature\cite{dashora1996temperature}.
Correspondingly, the material's specific heat capacity increases at elevated temperatures, primarily attributable to heightened kinetic energy manifested from atomic motion and augmented potential energy associated with interatomic bond distortions. The specific heat capacity for unaged specimens, similar to thermal conductivity, exhibits an abrupt shift in behavior at approximately 60\textdegree C, attributed to the melting of short-ordered crystallites. At the same time, it demonstrates a notable upward shift for all the aged specimens. When assessing the thermal conductivity and specific heat capacity of a 90-day moisture-aged sample compared to a 160-day moisture-aged sample, no significant variations were observed that warrant substantial commentary. However, a slight decline was noted in thermal conductivity and specific heat capacity upon comparing the unaged sample with the aged samples of the same material type. This reduction may be attributed to increased phonon scattering resulting from a higher inhomogeneous distribution of hard and soft domains and increased interphase resistance in the 90-day and 160-day aged samples.

\subsection{Influence of moisture and GMB volume fraction on Dielectric properties of syntactic foam}\label{dielectric_characterization}
\begin{figure}[H]
\centering
\subfigure[]{
\includegraphics[width=7.25cm]{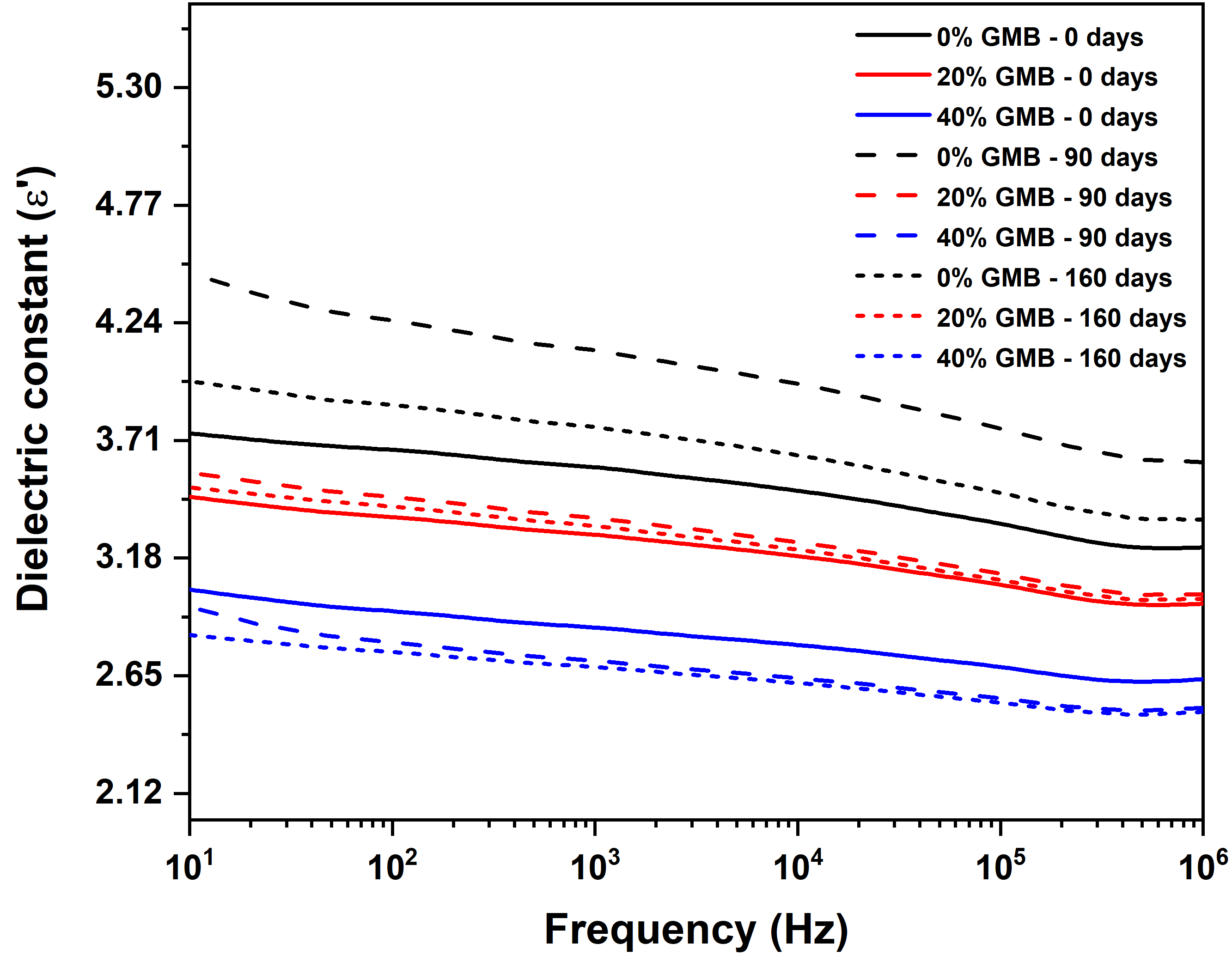}\label{Dielec_const}
}
\hspace{0.0in}
\centering
\subfigure[]{
\includegraphics[width=7.25cm]{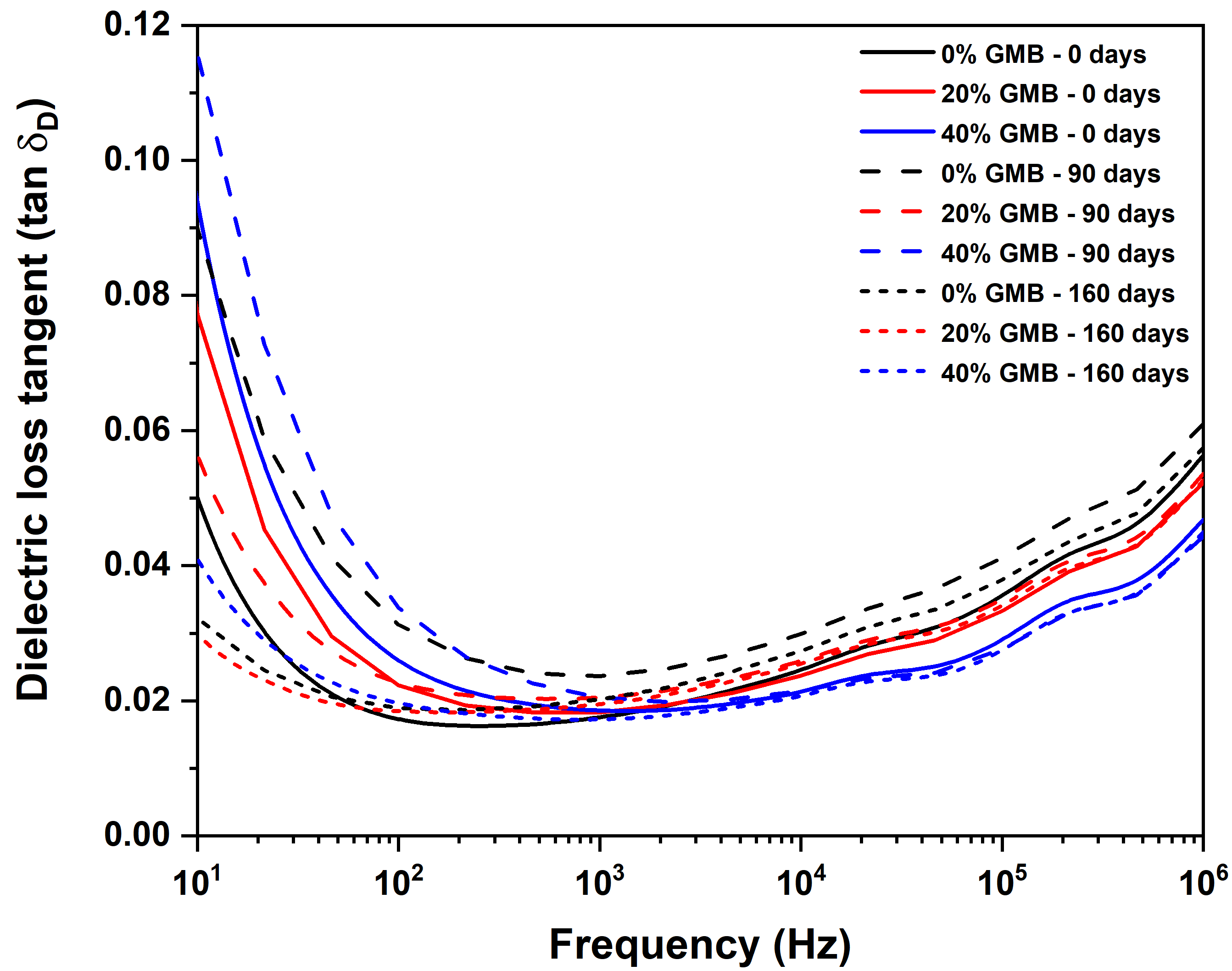}\label{Dielec_tandelta}
}
\caption{ Moisture driven changes dielectric properties as a function of GMB volume fraction and frequency (a) Dielectric constant and (b) Dielectric loss tangent}
\end{figure}

The dielectric constant explains the material's ability to store electrical charge through different types of polarization mechanisms (i.e., electronic, ionic, atomic, orientational, and interfacial) when we apply an electric field, that is achieved by the chemical compounds in polymers. The data depicted in Fig~\ref{Dielec_const} demonstrates that increasing the volume fraction of glass microballoons (GMB) reduces the material's dielectric constant. This phenomenon can be ascribed to the decline in the number of polymer chains participating in the polarization mechanism as the quantity of GMB increases. It is important to note that GMBs are hollow particles having a thin soda lime borosilicate glass covering, and it is commonly accepted that the dielectric constant of air is 1. A similar trend is also seen by other researchers\cite{Zhu,Zhu2014,Gupta2006}. The frequency domain dielectric constant response in all the samples shows a similar trend, as observed in Fig~\ref{Dielec_const}. A higher dielectric constant is evident in the lower frequency range $(< 100 Hz)$. This behavior is attributed to Maxwell-Wagner-Sillars polarization or interfacial polarization in all heterogeneous dielectric materials \cite{maxwell1873treatise,wagner1914after,sillars1937properties}. TPU, in this study, is a block copolymer with a multi-phase system where interfacial polarization occurs between hard and soft segments due to different dielectric constants, as reported in \cite{Jomaa2015, Petcharoen2013}. The orientational polarization phenomenon also contributes to the higher dielectric constant in the lower frequency range (10-1000 Hz). In this range, the soft and hard segments have sufficient time to orient themselves in response to an applied alternating electric field. This leads to easy mobility of molecules for orientational polarization, resulting in a high dielectric constant, as illustrated in Fig~\ref{Dielec_const}.

During the comparison of the change in dielectric constant for all samples at varying moisture exposure times, it was observed that the neat aged TPU showed a significant increase in dielectric constant across measured frequencies compared to unaged samples. However, after 160 days of moisture exposure, the dielectric constant decreased compared to 90 days but was still slightly higher compared to unaged TPU. A similar trend was noticed in the TPU with 20\% GMB. However, in the case of TPU with 40\% GMB, the dielectric constant decreased with an increase in the exposure time. One significant factor contributing to the observed trend is that neat TPU tends to have a higher concentration of chemical compounds that can participate in polarization. After 90 days of moisture exposure, distinct morphologies of hard segments are formed, as discussed in section~\ref{DSC-thermal_characterization}. These hard segments are highly polar (-NH-C=O) in nature and can easily undergo orientational polarization when an alternating field is applied. A similar trend is also observed in TPU with 20\% GMB. However, for TPU with 40\% GMB, the dielectric constant decreases after 90 days of exposure. This is because the GMB density per unit volume is higher in this specimen, which hinders the orientational polarization of the hard segments in the applied alternating electric field.

The dissipation factor, also called the loss tangent or tan delta, serves as a metric for quantifying energy loss or dissipation in dielectric materials under the influence of an alternating field. This dimensionless quantity is defined as the ratio of the imaginary part of permittivity to the real part of permittivity, and it measures the efficiency of dielectric materials without significant losses. A lower dissipation factor denotes a higher efficiency and superior performance of dielectric materials. However, several factors, such as molecular interactions, defects, and electrode interfaces, impede dipole alignment, which takes time to occur. Consequently, there is a delay between dipole orientation and the applied alternating field, leading to energy losses in the form of heat.

Fig ~\ref{Dielec_tandelta} shows the dielectric loss tangent (dissipation factor) for unaged and aged samples with different GMB volume fractions. It can be seen from the figure that the dissipation factor of unaged samples is higher for a higher volume fraction of GMB at 10 Hz frequency, which is attributed to interfacial polarization. Interfaces are formed when a filler or multiphase material is put together. The interfacial polarization is prominent at lower frequencies, resulting in a higher loss due to interface friction or interfacial polarization. The interfacial polarization becomes less dominant at higher frequencies, resulting in less loss and a lower dissipation factor. However, after $10^3$ Hz, there is a reversal in trend, with the dissipation factor being highest for neat TPU, followed by TPU with 20\% and 40\% GMB. This trend continues up to $10^6$ Hz. All specimens exhibit a slight increase in dissipation factor with increasing frequency beyond $10^3$ Hz. At higher frequencies, the dissipation factor reaches its highest value for the neat TPU sample, followed by TPU samples containing 20\% and 40\% GMB, respectively. The primary factor contributing to this behavior is the occurrence of leakage loss. It is important to note that a combination of relaxation polarization loss and leakage loss influences the dissipation factor\cite{He2020}. Consequently, an increase in the material's electrical conductivity leads to an escalation in leakage loss. The impact of leakage loss becomes more pronounced at higher frequencies due to the rapid alteration in the electric field, which prevents the molecules from aligning and adequately accommodating these charges, resulting in subsequent leakage.

After a 90-day exposure to moisture, it is observed that the dissipation factor of TPU containing 40\% GMB surpasses that of neat TPU and TPU with 20\% GMB at a frequency of 10 Hz. However, at a frequency of $10^3$ Hz, the dissipation factor of neat TPU exceeds that of TPU with 20\% and 40\% GMB. This trend persists in samples exposed to moisture for 160 days as well. The primary factors contributing to these observations are the combined effects of leakage and interfacial polarization loss. Specifically, at 10 Hz and after 90 days of moisture exposure, the TPU exhibits higher electrical conductivity, followed by TPU with 40\% GMB and TPU with 20\% GMB. Simultaneously, a higher volume fraction of GMB enhances interfacial polarization loss.
Consequently, TPU with 40\% GMB demonstrates a higher dissipation factor, followed by neat TPU and TPU with 20\% GMB. A similar trend is observed after 160 days of moisture exposure, albeit with significantly lower dissipation factor values. This can primarily be attributed to the confinement of highly polar hard segments within the soft domain, preventing the polarization of molecules under the applied electric field.

%%%%%%%%%%%%%%%%%%%%%%%%%%%%%%%%%%%%%%%%%%%%%%%%%%%%%%%%%%%%%%%%%%%%%%%%%%%%%%%%%%%%%%%%%%%%%%%%%%%%%%%%%

\section{Conclusion}\label{conc}
\begin{figure}[h!]
  \includegraphics[width=6.25in]{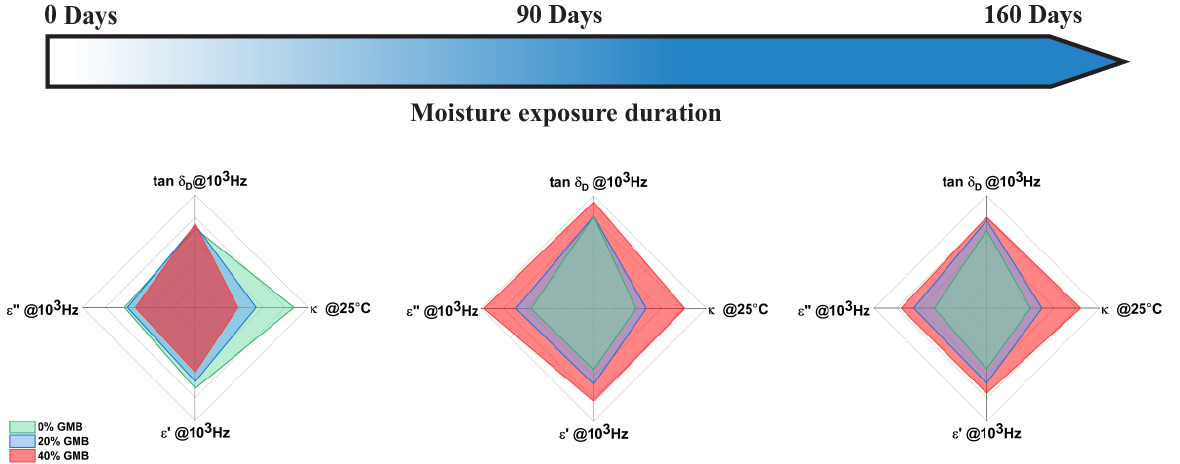}
  \caption{Summary of thermal and dielectric property evolution with exposure times}\label{fig:summary_results}
\end{figure}

This paper aims to explore the long-term moisture-induced degradation in TPU and TPU-based syntactic foams, along with their associated microphase morphology changes. To this end, samples were immersed in water and tested at two different time points: 90 days and 160 days. Additionally, the study aims to shed light on the manifestation of these changes in thermal and dielectric properties, and the maps of how these properties evolve are summarized in Fig~\ref{fig:summary_results}.

%The research presents the following key insights:
%\begin{enumerate}

\paragraph{Thermal Transport}
The TPU-based syntactic foam's thermal conductivity and specific heat capacity decrease as the volume fraction of GMB increases from 0\% GMB, 20\% GMB, and 40\%. For each of these volume fractions, the thermal conductivity remains relatively stable with temperature variations up to 100\textdegree C, except for a sharp increase around 60°C, corresponding to the melting of short-ordered crystallites. Similarly, the specific heat capacity generally increases with temperature but exhibits a noticeable change around 60\textdegree C. In the case of moisture-exposed samples, this change shifts to higher temperatures due to moisture-induced dissociation and reorganization of hard segments. At room temperature (25°C), it was observed that the thermal conductivity of 40\% GMB reinforced syntactic foam exposed to moisture for 90 days had a negigible change compared to their unaged values. Further, the difference in thermal conductivity for all foams between the 90-day and 160-day moisture exposure was negligible.

\paragraph{Dielectric Properties}
The dielectric constant decreases as the volume fraction of GMB increases. Additionally, it follows a decreasing trend as the frequency of the applied field increases. This can be attributed to the limited time available for the molecules to orient themselves in response to the applied field, consequently reducing the material's dielectric constant.
It is important to note that after 160 days of moisture exposure, there is a 35.4\% reduction in the dielectric loss factor of neat TPU at 10 Hz compared to unaged neat TPU, and a 1.86\% reduction at $10^6$ Hz. However, there was a 79.5\% increase in the dielectric loss factor in neat TPU exposed to moisture for 90 days at 10 Hz, whereas, an 8.20\% increase in the dielectric loss factor at $10^6$ Hz. It is noteworthy that similar behavior is also observed in TPU-based syntactic foams with 20\% and 40\% GMB when compared to their unaged counterparts. One possible reason is the distribution morphology of the hard and soft segments in TPU(GMB's are unaffected by moisture). At 90 days, the disordered hard domains are more prevalent, whereas after 160 days of moisture exposure, the hard segments are uniformly distributed, and the hard domain content is decreased.

The investigation of the thermal transport and dielectric behavior of TPU-based syntactic foam serves a dual purpose: it enhances our understanding of the moisture-induced degradation mechanisms and provides valuable insights into the geometrical hierarchy at various length scales that can be customized to attain desired properties. For instance, at the nanoscale level, the microphase morphology exhibits increased disordered hard domains, leading to an increase in the dielectric constant and dielectric loss. Similarly, when the soft segment length is reduced, it promotes the homogeneous distribution of hard segments and reduces the hard domain region, resulting in a decrease in dielectric loss. At the microscale level, the GMB content can be adjusted to modify the void content, which directly impacts the viscoelastic\cite{SUBRAMANIYAN2023110547}, thermal transport, and dielectric properties. The additive manufacturing technique used in this study offers an added advantage, as it enables the fabrication of tunable materials at the macroscale level.

%\end{enumerate} 

%%%%%%%%%%%%%%%%%%%%%%%%%%%%%%%%%%%%%%%%%%%%%%%%%%%%%%%%%%%%%%%%%%%%%%%%%%%%%%%%%%%%%%%%%%%%%%%%%%%%%%%%%

\section*{Author Contributions}

{\bf Sabarinathan P Subramaniyan:} Conceptualization, Methodology, Formal Analysis, Visualization, Investigation, Writing - Original Draft. {\bf Partha Pratim Das:} Methodology, Investigation, Writing- Reviewing and Editing. {\bf Rassel Raihan:} Methodology, Investigation, Writing- Reviewing and Editing. {\bf Pavana Prabhakar:} Conceptualization, Methodology, Writing - Original Draft, Visualization, Verification, Supervision, Project Administration, Funding Acquisition. 
%%%%%%%%%%%%%%%%%%%%%%%%%%%%%%%%%%%%%%%%%%

%%%%%%%%%%%%%%%%%%%%%%%%%%%%%%%%%%%%%%%%%%

\section*{Funding}

The authors would like to acknowledge the support of the National Science Foundation (NSF) CAREER Award \# 2046476 through the Mechanics of Materials and Structures (MOMS) Program for conducting the research presented here.

%%%%%%%%%%%%%%%%%%%%%%%%%%%%%%%%%%%%%%%%%%

\section*{Acknowledgment}
This research was partially supported by the University of Wisconsin - Madison College of Engineering Shared
Research Facilities and the NSF through the Materials Science Research and Engineering Center (DMR-1720415)
using instrumentation provided at the UW - Madison Materials Science Center. The authors express their gratitude to the Polymer Engineering Center for providing access to the Netzsch LFA 447 instrument to conduct Laser Flash Analysis.
 
%%%%%%%%%%%%%%%%%%%%%%%%%%%%%%%%%%%%%%%%%%

%\section*{References}

{
\typeout{}
\bibliographystyle{ieeetr}
%\biboptions{numbers, square, compress}
\bibliography{references}
}

%----------------------------------------------------------------------------------------
\section*{Appendices}
\subsection*{Dielectric study}
\begin{figure}[H]
\centering
\subfigure[]{
\includegraphics[width=7cm]{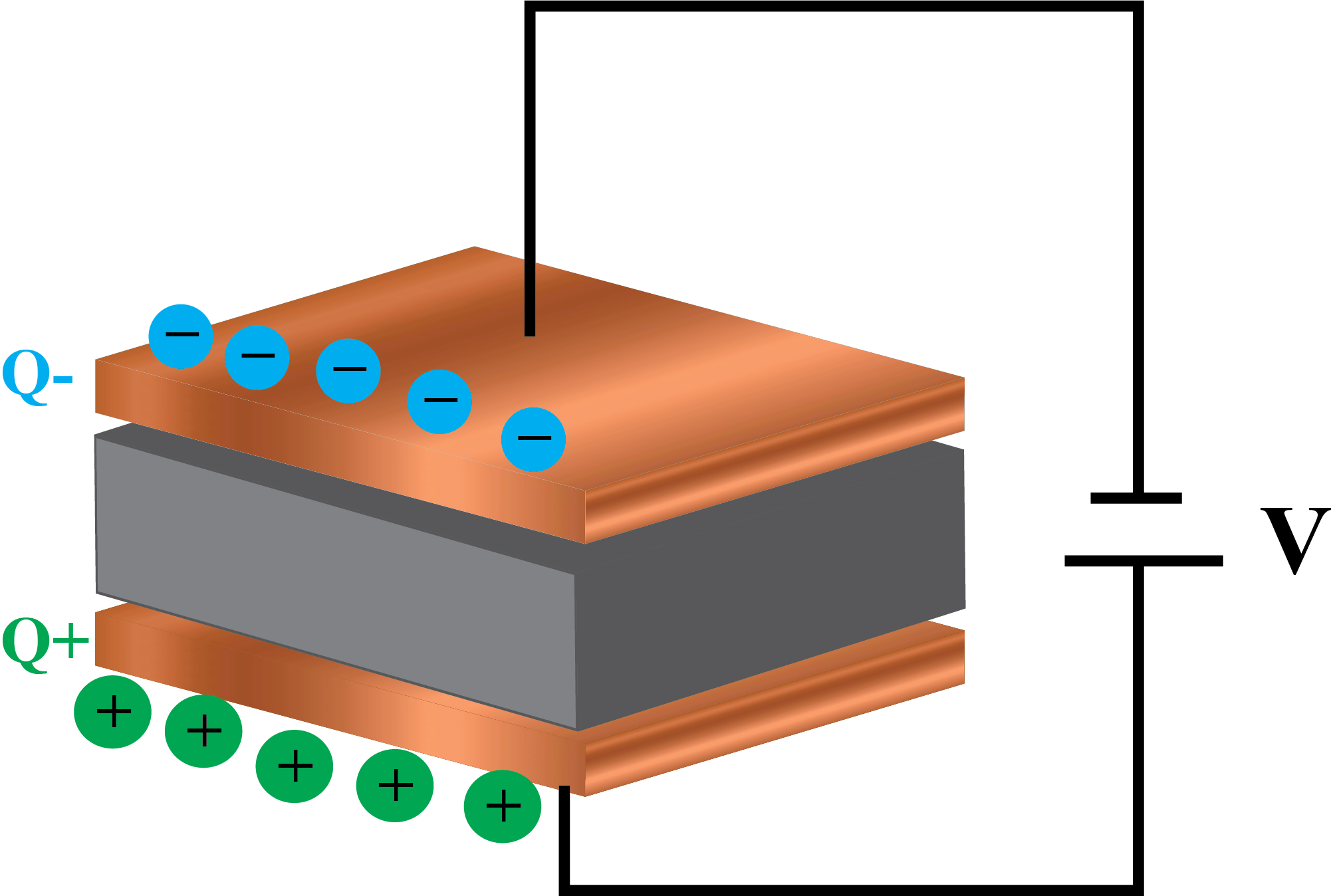}
}
\hspace{0.2in}
\centering
\subfigure[]{
\includegraphics[width=8
cm]{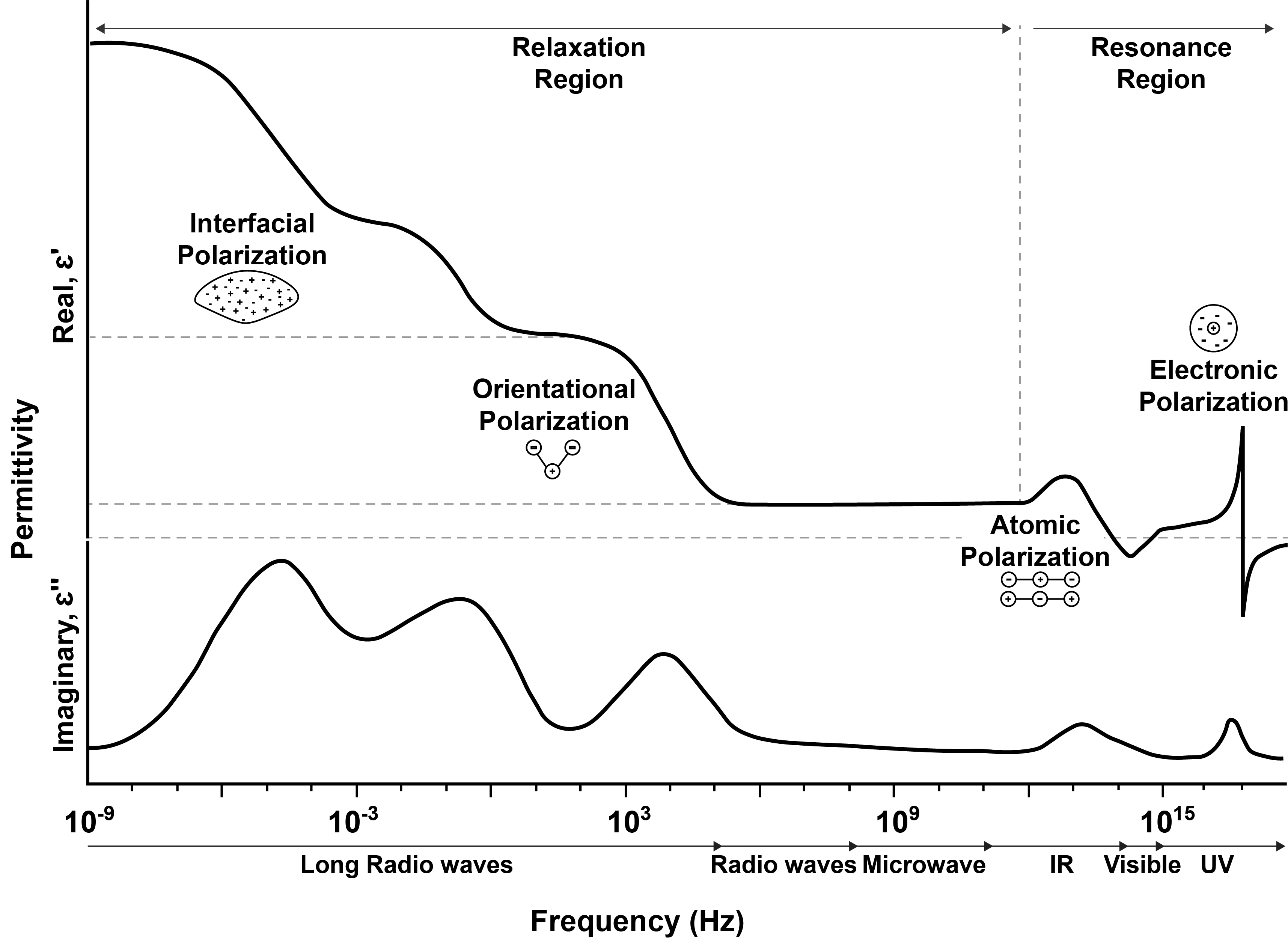} 
}
\caption{(a) Schematic illustration of dielectric analysis and (b) Dielectric response and attributing mechanism at different frequencies\cite{das2022dielectric}}\label{dielectric_mechanism}
\end{figure}

The dielectric mechanisms occurring across a broad spectrum of frequencies are depicted in Fig ~\ref{dielectric_mechanism}. In the present study, the analyzed frequency spectrum manifests two distinct polarization phenomena. The first pertains to space charge or interfacial polarization, while the second corresponds to orientational or dipolar polarization. Conversely, atomic and electronic polarization phenomena are observed at significantly higher frequencies, as visually represented in the Fig~\ref{dielectric_mechanism}.

\textbf{Orientation or dipolar polarization:} Polymers are composed of various molecules that arise from the combination of atoms, resulting in the sharing of electrons. This sharing of electrons leads to an uneven charge distribution, resulting in permanent dipole moments that are randomly arranged in the absence of an electric field. However, applying an electric field results in torque exertion on the dipole, causing it to rotate and eventually align with the electric field. This alignment results in orientational polarization, which, in turn, leads to the occurrence of friction and dielectric losses\cite{note2006agilent}.

\textbf{Space charge or orientational polarization:} If a low-frequency electric field is applied, charge carriers may traverse through the material over a given distance. This movement of migrating charges can be obstructed, leading to interfacial or space charge polarization. The charges may be confined to the interfaces of a material if they are unable to be freely discharged or replaced at the electrodes. This results in a distortion of the electric field, which leads to an increase in the material's overall capacitance and, consequently, an increase in the dielectric constant. At low frequencies, the charges have sufficient time to accumulate at the borders of the conducting regions, leading to an increase in $\epsilon'$. However, at higher frequencies, the charge displacement is relatively small compared to the dimensions of the conducting region, and therefore, polarization does not occur\cite{note2006agilent}.

In the context of dielectric measurement principles, the experimental setup involves positioning the sample within a simple capacitor formed by two electrodes. Subsequently, a voltage denoted as $U_0$, operating at a fixed frequency of $\omega$/2$\pi$ is applied to the sample, resulting in a corresponding current Io at the identical frequency. Notably, the system response exhibits a phase shift or phase lag, characterized by the phase angle $\phi$, representing the temporal difference between the voltage and current waveforms. $U^*$,$I^*$,$Z^*$, $\epsilon^*$, and $tan \phi$ denote the complex voltage, complex current, complex impedance, complex permittivity, and dielectric loss factor, respectively. 
\begin{equation}
\begin{split}
    U(t)  = U_0 cos(\omega t) \\
    I(t) = I_0 cos(\omega t + \phi)
\end{split}
\end{equation}

\begin{equation}
\begin{split}
    U^* = U'+iU''\\
    I^* = I'+iI''
    \end{split}
\end{equation}

\begin{equation}
\begin{split}
    U' = U_0\\
    U'' = 0\\
    I' = I_0 cos\phi\\
    I'' = I_0 sin\phi
\end{split}
\end{equation}

\begin{equation}\label{dielectric_tan_equation}
    tan\phi = \frac{I''}{I'}
\end{equation}

\begin{equation}\label{Impedence}
    Z^* = \frac{U^*}{I^*}
\end{equation}

Dielectric properties of the material are related to the impedance measured using Equation \ref{Impedence}. Where $C_0$ is the capacitance of the empty sample capacitor.
\begin{equation}
    \epsilon^* = \epsilon' - i\epsilon'' = \frac{-i}{\omega Z^*(\omega)} \frac{1}{C_0}
\end{equation}

\begin{figure}[H]
 \centering
  \includegraphics[width=8cm]{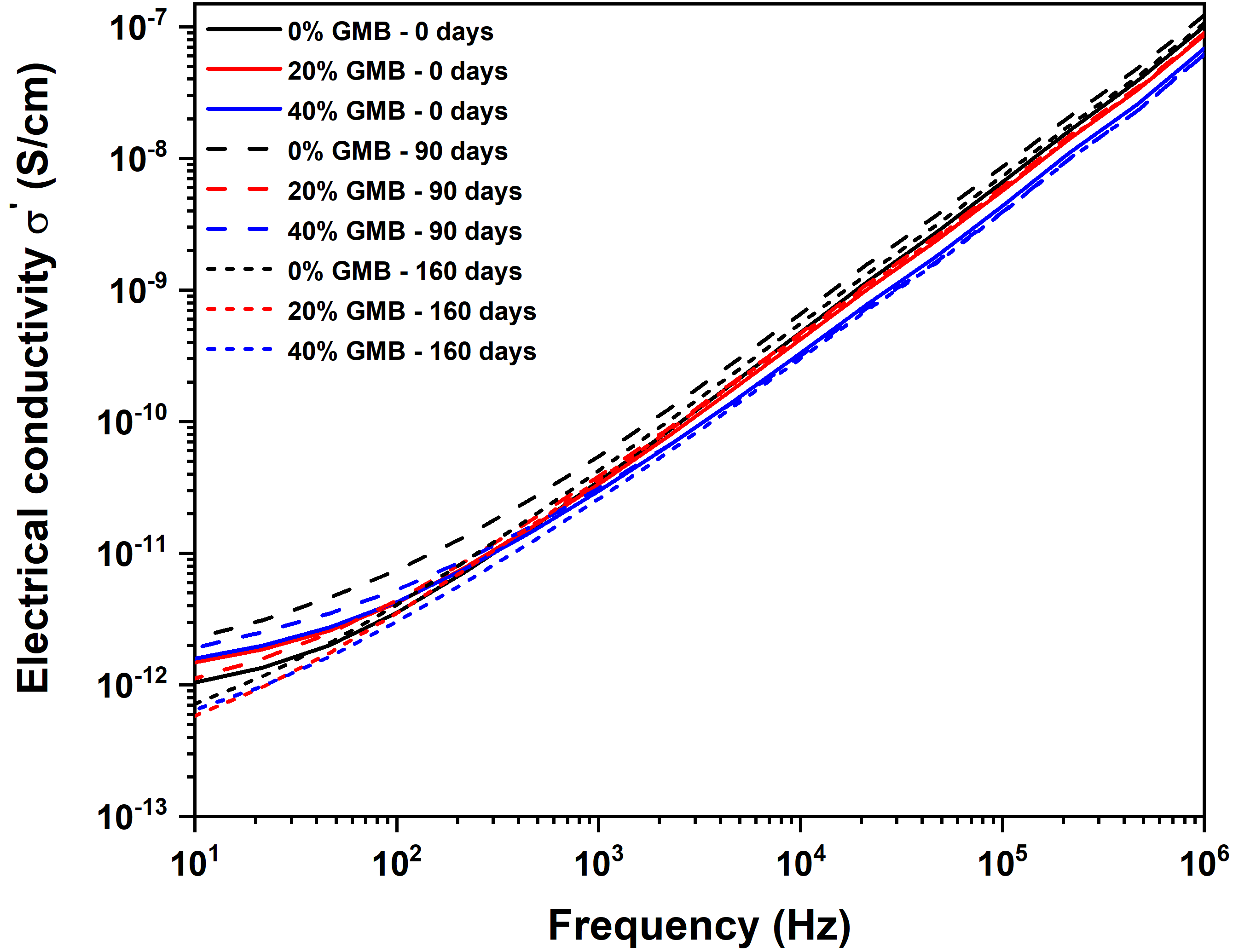}
  \caption{Electrical conductivity of TPU and TPU based syntactic foam as a function of moisture aging time and frequency}\label{electrical_cond}
\end{figure}

\end{document}